# Preparation of clean surfaces and Se vacancy formation in $Bi_2Se_3$ by ion bombardment and annealing

Weimin Zhou, Haoshan Zhu, Connie M. Valles and Jory A. Yarmoff*

*Department of Physics and Astronomy, University of California, Riverside CA 92521*

**Abstract**

Bismuth Selenide ($Bi_2Se_3$) is a topological insulator (TI) with a structure consisting of stacked quintuple layers. Single crystal surfaces are commonly prepared by mechanical cleaving. This work explores the use of low energy $Ar^+$ ion bombardment and annealing (IBA) as an alternative method to produce reproducible and stable $Bi_2Se_3$ surfaces under ultra-high vacuum (UHV). It is found that a well-ordered surface can be prepared by a single cycle of 1 keV $Ar^+$ ion bombardment and 30 min of annealing. Low energy electron diffraction (LEED) and detailed low energy ion scattering (LEIS) measurements show no differences between IBA-prepared surfaces and those prepared by *in situ* cleaving in UHV. Analysis of the LEED patterns shows that the optimal annealing temperature is 450°C. Angular LEIS scans reveal the formation of surface Se vacancies when the annealing temperature exceeds 520°C.

---

*Corresponding author, E-mail: yarmoff@ucr.edu

## I. Introduction

Bi$_2$Se$_3$ has been studied for decades as a classic thermoelectric material [1,2] and recently has drawn more attention as a second generation three-dimensional (3D) topological insulator (TI) with a practical band gap around 0.3 eV [3,4]. The topological surface states (TSS) of a TI material make its boundaries conductive while the bulk remains insulating. This unique property makes it promising for use in applications such as topological superconductors [5], quantum computers [6] and spintronics-based devices [7,8].

Bi$_2$Se$_3$ is a layered material consisting of stacked quintuple layers (QL) that are held together by weak van der Waals forces [9]. Although this enables ordered surfaces to be prepared by mechanical cleaving along the (001) plane, cleaving has many disadvantages. For example, *in situ* cleaving under ultra-high vacuum (UHV) requires special preparation by attaching parts to each individual sample [10] and installation of tools inside the chamber [11]. More importantly, the *in situ* procedure cannot guarantee an optimal cleave, as sometimes a flat surface is obtained only after several consecutive cleaves from a particular sample. *Ex situ* cleaving is even more problematic as the resulting surface structures have been found to vary from sample to sample [11-15]. For example, some studies indicate a Se-termination [15], as expected from the bulk crystal structure, while others suggest that the bulk-terminated surface is covered by an extra bilayer of Bi [13,16].

Ion bombardment and annealing (IBA) is a straightforward and powerful UHV surface preparation method that has been utilized for more than 60 years since the pioneering work of Farnsworth *et al.* [17]. IBA involves the removal of surface contaminants by sputtering with low energy inert noble gas ions followed by recrystallization of the damaged surface by annealing in vacuum.

There have been some previous investigations in which IBA was used to prepare $Bi_2(Se,Te)_3$ materials. Most of these employed low energy electron diffraction (LEED) or scanning tunneling microscopy (STM), although neither of these can directly provide the surface composition or local structure below the top atomic layer. Haneman used LEED to show that $Bi_2Te_3$ surfaces prepared by *in situ* cleaving and IBA using annealing temperatures in the range of 200°C to 260°C have similar symmetries, but the detailed surface structure was not investigated [18]. STM images by Coelho *et al.* show that IBA-prepared $Bi_2Te_3$ using an annealing temperature of 300°C is approximately terminated half by Te and half by a Bi bilayer [19]. These terminations were inferred from STM line-profiles using the fact that the thickness of a Bi bilayer is about half that of a $Bi_2Te_3$ QL. Cavallin *et al.* used STM to show that a Bi-bilayer film appears on IBA-prepared $Bi_2Se_3$ surfaces when annealed at around 320°C [20].

A previous study from our group used low energy ion scattering (LEIS) to compare surfaces prepared by *in situ* cleaving, *ex situ* cleaving and IBA [11]. In contrast to STM or LEED, LEIS directly reveals the surface composition and structure of the outermost two to three atomic layers [21]. It was found that *in situ* cleaved and IBA-prepared surfaces are Se-terminated and the surface atomic structures are very similar, but *ex situ* cleaved surfaces can either have a Se-termination or be Bi-rich [11]. XPS and LEIS showed the presence of surface contamination on *ex situ* cleaved surfaces, and it was inferred that a chemical reaction with the contaminants contributes to the non-reproducibility of their termination.

In this paper, surface preparation of $Bi_2Se_3$ by IBA is investigated in more detail using LEIS and LEED. It is shown that a well-ordered Se-terminated surface can be successfully prepared by IBA using an optimized annealing temperature, and that such a surface has excellent stability even after exposure to air. In addition, impact collision ion scattering spectroscopy (ICISS), an advanced variation of LEIS, reveals the formation of Se vacancies in the near-surface region after annealing at higher temperatures. Such vacancies are not observable by LEED alone. The ability to experimentally detect Se vacancies is of great importance as they are believed to be responsible for the natural n-type doping of $Bi_2Se_3$ that mixes the TSS with the bulk conduction states [1,22,23].

## II. Experimental Procedure

Single crystals of $Bi_2Se_{3.12}$ were grown by melting stoichiometric mixtures of Bi and Se shot in an evacuated quartz ampule and then following a slow-cooling protocol [23,24]. The details of the sample growth recipe and surface preparation methods are provided elsewhere [11], so only the basic information is given here. The samples were synthesized with a slight enrichment of Se to compensate for the possibility of bulk Se vacancies, but such a stoichiometry adjustment does not affect the surface properties [11]. After growth, the ingots are fragmented with a razor blade to obtain flat and shiny pieces. These pieces can be easily cleaved along the (001) plane by conventional exfoliation methods to produce a surface suitable for investigation. A typical sample diameter is around 10 mm.

Atomic force microscopy (AFM) measurements are conducted in tapping mode using a Dimension 5000 AFM (Digital Instruments) with TESPA-V2 silicon tips (Bruker). The samples prepared by IBA are removed from UHV and exposed to air during the AFM measurements.

Ar$^+$ ion bombardment, LEED and LEIS are all carried out in a UHV chamber with a base pressure of $2 \times 10^{-10}$ Torr. The UHV chamber is attached to a load lock that enables quick introduction of new samples without the need for baking the main chamber. The samples are attached to transferable Ta sample holders (Thermionics) that have a diameter of 30 mm and a thickness of 1 mm. The holders are fitted in UHV onto the foot of a vertical x-y-z manipulator that has two rotational degrees of freedom.

The foot of the manipulator contains a filament mounted behind the sample holder that includes the ability to operate as an electron beam heater by biasing the filament with a high negative voltage relative to the sample. For the work performed here, however, the samples are simply heated radiatively by running current through the filament without applying a bias voltage. A calibration of the sample temperature to the filament current was performed by temporarily attaching several thermocouples to an empty sample holder and monitoring the temperature as a function of filament current. Thermocouple measurements showed that the temperature in the center of the sample holder is about 20°C hotter than it is 5 mm from the center and about 100°C hotter than it is 10 mm away, but it is the temperature at the center of the sample holder that is reported here. Also, as the calibration was done with no sample attached to the holder, the temperatures reported could be slightly larger than the actual temperature of the $Bi_2Se_3$ samples. Finally, note that it takes more than half an hour for the sample holder temperature to stabilize and the thickness of the sample also affects the surface temperature. An infrared pyrometer, which can detect temperatures above about 300°C, is also used during annealing to roughly monitor temperature changes over time. These issues with the temperature measurement lead to a larger uncertainty that is the sample may be cooler than reported, rather than warmer, and are thus estimated to be about -50°C, +20°C. Despite concerns about the absolute accuracy of the temperature measurements, however, they correctly provide relative values when compared to each other.

Samples that are to be prepared by IBA are mounted onto the transferable sample holder using spot welded Ta strips. The samples are cleaved in air by carbon tape multiple times to obtain a visibly flat surface. Note that some of these *ex situ* cleaved samples are measured by AFM without any additional surface treatment. The samples to be subjected to IBA are inserted into the main UHV chamber and subjected to multiple cycles of $Ar^+$ ion bombardment using a backfilled sputter ion gun (Perkin-Elmer) and annealing. Ion energies employed for sputtering are in the range of 0.5 to 1.0 keV and the background pressure of Ar is $2\times10^{-5}$ Torr. The size of the sputter ion beam is defocused to be larger than the sample to minimize any non-uniformity in the flux. The surface atomic density of $Bi_2Se_3$ is about $6.8\times10^{14}$ $cm^{-2}$. The average beam flux is about 200 nA $cm^{-2}$ as measured by a Faraday cup, so that after 30 min of bombardment the ion fluence will be $2.3\times10^{15}$ $cm^{-2}$, which is sufficient to remove the top few atomic layers. After sputtering, the samples are annealed in UHV to desorb the embedded Ar and recrystallize the surfaces.

Samples to be cleaved *in situ* are mounted onto the transferable holder with Ag paste, while a small Al bar is attached to the front of the sample with Torr-Seal, as described in Refs. [10,11]. The Al bar is knocked off inside the UHV chamber to cleave the sample and obtain a flat and shiny surface.

A Na$^+$ thermionic emission ion gun (Kimball Physics) is used for the LEIS measurements. For the experiments performed here, the incident ion beam energy is fixed at 3.0 keV. The ion gun is mounted on a turntable that can rotate around the vertical axis of the chamber. The diameter of the focused ion beam is less than 1 mm. The manipulator allows the samples to rotate azimuthally around the surface normal and about the vertical axis that is in the plane of the sample's surface. In this way, the scattering angle, the polar angle and the azimuthal angle can each be changed independently. The polar angles are reported with respect to the plane of the sample surface. A stepper motor is controlled by a computer to automatically adjust the polar angle. The chamber contains two detectors for LEIS, a straight tube for time-of-flight (TOF) and a hemispherical electrostatic analyzer (ESA). The incident ion fluence for all LEIS data is kept below 1% of a monolayer to limit any surface damage.

TOF uses a pulsed ion beam and measures the flight time of the scattered ionic and neutral projectiles, as previously described [25]. For the work performed here, only the total scattered yield is presented. The Na$^+$ ion beam is pulsed at 80 to 100 kHz by using deflection plates to pass the beam across a 1 mm aperture mounted in front of the ion gun. The scattered particles are detected by a triple microchannelplate (MCP) array mounted at the end of a 0.57 m long flight tube at a scattering angle of 125°. The entrance to the MCP is held at ground potential so that the sensitivity to neutral and ionic species is the same. There are two 3 mm diameter apertures in the flight tube, which limits the acceptance angle to less than 1°.

Energy spectra and of the scattered ion yield and ICISS angular distributions are measured using a 160° Comstock AC-901 hemispherical ESA that has a radius of 47.6 mm and 2 mm diameter entrance and exit apertures, making the acceptance angle approximately 2°. The scattering angle used with the ESA is 161°.

ICISS simulations were performed with Kalypso, a software package that uses molecular dynamics (MD) to simulate atomic collisions in solids [26]. The recoil interaction approximation (RIA) is utilized so that only the repulsive interactions between the projectile and individual target atoms need to be considered. The Thomas-Fermi-Moliére repulsive screening potential is used with the Firsov screening length corrected by a factor of 0.8 and a potential cut-off distance of 2.9 Å [27]. The structural parameters of the Se-terminated simulation target is the average of the two sets of experimental parameters obtained via LEED and SXRD [28] with the spacing between the first and second atomic layer set to 1.55 Å as determined in Ref. [27]. The vibrational amplitudes of the atoms in the bulk crystal are established using the bulk Debye temperature of 200 K [29]. The mean square vibrational amplitudes of the top two atomic layers are isotropically enhanced by a factor of 2 by setting their Debye temperature to 140 K. The acceptance range of the scattering angles for the simulations is chosen to be 2° to match that of the experimental data.

**III. Results and Discussion**

LEED patterns provide the symmetry of the two-dimensional surface unit cell and can be used as a measure of the crystallinity in the near-surface region [30]. In the present work, they are also used to identify the appropriate annealing temperature and the minimum number of cycles needed for IBA to produce a highly ordered surface. The LEED pattern collected from a sample that is cleaved *ex situ* before being subjected to the IBA process is very dim indicating a disordered surface due to atmospheric contaminants [11].

After insertion into the UHV main chamber, the samples are first degassed at 130°C for several hours to remove adsorbed water and hydrocarbons, followed by two-hours of $Ar^+$ sputtering and then a 30 min anneal at 130°C. The samples are next subjected to successive IBA cycles using higher annealing temperatures, with LEED patterns collected after each cycle. One IBA cycle is defined here as 30 min of 0.5 to 1.0 keV $Ar^+$ bombardment and 30 min of annealing.

Figure 1(a) shows a representative LEED pattern collected from a $Bi_2Se_3$(001) surface after a 30 min 1.0 keV $Ar^+$ ion bombardment and a 30 min anneal at 510°C. Figure 1(b) shows the gray scale line profile along the line indicated in Fig. 1(a) determined with the ImageJ software package. The full width at half maximum (FWHM) of the peaks in Fig. 1(b) are used to represent the diameters of the LEED spots. The FWHM's are then used to calculate the average domain sizes of the surface assuming that domain size is the only factor that contributes to the width of the LEED spots, i.e., other contributions to the spot size, such as the diameter of the electron beam and aberrations in the LEED optics, are ignored. Although such domain sizes are not accurate, relative changes to their sizes as a function of sample treatment do provide useful information.

Figure 2 shows the calculated domain sizes as a function of annealing temperature. For ion bombarded samples, a blurry LEED pattern with large spots is always visible prior to annealing even after hours of sputtering as indicated by the first point in Fig. 2. After annealing, the domain size increases until it reaches a maximum at around 450°C after which it starts to decline, suggesting that 450°C is the optimal annealing temperature. The domain size after a 450°C anneal is comparable to that of an *in situ* cleaved surface, which is also indicated in Fig. 2. Once a particular sample has been treated with the preliminary degassing and low temperature annealing procedure, a single IBA cycle is sufficient to achieve a sharp LEED pattern.

AFM shows that IBA produces surfaces with uniform steps and flat terraces. Figure 3 shows typical AFM images collected from an IBA-prepared surface annealed at 490°C. Figure 3(a) shows the image in a 3D, oblique perspective that acts to highlight the edges of the steps. Many step edges and large triangles of a size on the order of μm are observed across the surface. Figure 3(b) is a two-dimensional (2D) representation of a selected area that uses a color scheme to indicate the height directly. Many flats terraces separated by step edges are seen in Fig. 3(b). Figure 3(c) is a line profile measured along the line indicated in Fig. 3(b). The heights of all of the steps in the line profile are around 1 nm, which is close to the 0.95 nm height of a bulk $Bi_2Se_3$ QL [12]. The terraces are 0.2 μm wide on average and the roughness on the terraces is typically around 0.5 Å, which is close to the resolution of the AFM instrument. This measured roughness is smaller than the typical 0.8 Å roughness that is measured on *ex situ* cleaved surfaces. *Ex situ* cleaved surfaces also show a much smaller density of step edges due to their increased roughness.

LEIS, a straightforward and powerful technique for surface analysis [21], is employed here to determine the surface composition and local atomic structure on the terraces. The incident kinetic energies in LEIS are on the order of keV, which are much larger than the surface bonding energies, so that interactions between target atoms can be ignored.

Furthermore, the scattering cross sections are smaller than interatomic spacings allowing for the use of the binary collision approximation (BCA), which assumes that the projectiles interact with only a single target atom at a time. Also, in the low energy ion range, the collisions can be considered to be completely classical as the de Broglie wavelength is on the order of a hundredth of an Å. If a projectile escapes the sample after only one collision, the kinetic energy of the scattered projectile will primarily depend on the projectile/target mass ratio and the scattering angle [21]. In this way, each surface element that is directly visible to both the ion beam and the detector contributes a single scattering peak (SSP) to a LEIS spectrum. The particles singly scattered from heavier target atoms will have a larger kinetic energy than those singly scattered from lighter elements.

The surface sensitivity in LEIS with alkali projectiles is largely due to shadowing and blocking effects [21]. A shadow cone is defined as the region behind a surface atom into which a projectile cannot penetrate because of the repulsive interatomic potential. A blocking cone is a similar concept, but applies to projectiles that are blocked by shallower atoms as they exit the surface after undergoing a large angle collision with a deeper lying atom. Because the size of the scattering cross sections is comparable to the interatomic spacings, atoms can effectively shadow or block projectiles from regions behind a target atom leading to the ability to define how many atomic layers a particular measurement will probe.

Figure 4 shows TOF spectra collected from a surface prepared by IBA with an annealing temperature of 510°C and from an *in situ* cleaved surface. In TOF-LEIS, the more energetic particles reach the detector more quickly, which is why the x-axis is plotted in reverse. Thus, the projectiles singly scattered from Bi will arrive at the detector earlier than those scattered from Se. The large backgrounds in Fig. 4 that extend from about 9.0 to about 4.5 µs are due to multiple scattering trajectories. The actual background should increase at longer flight times as a consequence of the cascade of multiply scattered particles, but the efficiency of the MCP detector goes down quickly at lower impact energies [31] leading to an apparent decrease in intensity at longer flight times.

A double alignment orientation is used for the spectra shown in Fig. 4 so that only the outermost atomic layer contributes to the SSP, as described in Ref. [11]. Double alignment implies that both the incident ion beam and the detector are positioned along low index crystalline directions. The use of normal incidence prevents the incident ions from reaching beyond the third atomic layer, as ions scattered from the first layer shadow those in the fourth layer, the second layer shadows the fifth and the third layer shadows the sixth. By placing the detector at the outgoing low index direction corresponding to the [210] azimuth at a polar angle of 36°, the ions scattered from the third layer are blocked by the second layer and those scattered from the second layer are blocked by the atoms in the first layer. Thus, in this orientation only single scattering events from the outermost layer are detected. The TOF-LEIS spectra in Fig. 4 show only one feature at 6.3 µs, which is identified as the Se SSP. The Bi SSP would appear at 4.8 µs, but its absence indicates that both the IBA-prepared and *in situ* cleaved surfaces are Se-terminated, in agreement with Ref. [11].

The angular dependence of the SSP intensities is sensitive to the detailed atomic structure of the outermost few layers. As the sample orientation is adjusted, scattered projectiles are deflected in and out of the shadow and blocking cones leading to changes in the scattered ion yield. Since the flux of projectiles is maximized at the edges of the shadow and blocking cones, the detailed shape of the angular distributions can be rather complex. When a large scattering angle is employed, the features in the angular scans are primarily due to shadowing effects. This experimental arrangement is known as ICISS [32,33].

Figure 5 shows a representative energy spectrum collected with the ESA using 3.0 keV Na$^+$ scattering from an IBA-prepared surface. The peak at 0.8 keV is the Se SSP while the peak at 1.8 keV is the Bi SSP. Both SSP's are visible as this is not a double alignment orientation. The background is considerably different than in the TOF spectra because the ESA only measures the yield of ions and, unlike the MCP detector used for TOF, its transmission function does not decrease with scattered energy. This spectrum is shown here as it is used to locate the energies of Se and Bi SSPs so that the ESA can be fixed at a particular SSP energy to collect an ICISS angular distribution.

Figure 6 shows ICISS scans of the Se SSP intensity as the incident polar angle is adjusted by the stepper motor along the [010] azimuth from surfaces prepared by IBA with annealing temperatures of 510°C and 520°C, along with the data from an *in situ* cleaved surface. The features in these scans are formed as the sample rotates such that flux at the edges of the shadow and blocking cones interacts with other atoms in the crystal from which the projectiles singly scatter.

When the incident beam is close to the plane of the surface, there will be no backscattered yield because all of the surface atoms are located within the shadow cones of their neighboring surface atoms. As the incident polar angle increases to a point called the critical angle, the edges of the shadow cones of the first layer will pass through their neighboring first layer atoms, which is illustrated in the upper drawing in the inset to Fig. 6.

As the ion beam is rotated, the intensity of the Se SSP reaches a maximum due to the increased flux at the shadow cone edges. Such a maximum due to interaction of a first layer shadow cone with its nearest neighboring first layer atom is called a surface flux peak (SFP). The peak at around 23° in Fig. 6 is the SFP, although in this case it results from a combination of the first and third Se layers since these are equivalent planes that are both visible to the beam and detector along the [010] azimuth, as discussed below.

As the polar angle increases beyond the SFP, interactions of the shadow cones formed around the surface atoms with deeper lying atoms lead to additional features in the ICISS polar scans. In this manner, every feature in an ICISS polar scan generally corresponds to a pair of interacting atoms in the solid created when the edge of the shadow cone of one atom passes through another atom in the crystal structure. By careful analysis, it is thus possible to identify each feature in an ICISS polar scan by considering all of the pairs of interacting atoms.

The $Bi_2Se_3$ structure along the [010] azimuth is a complicated case, however, as there are three inequivalent (300) planes visible to the ion beam and detector that are particularly close to each other, as illustrated in Fig. 7. Two of the (300) planes indicated by the vertical strips (right and left) are terminated with Se in the first and third layers, respectively, while the other (300) plane (middle) is terminated with Bi in the second layer. The distance between the (300) planes is only 1.2 Å, which is smaller than the radius of the shadow and blocking cones arising from atoms in the top two layers. Thus, ions singly scattered from the third layer atoms not only interact with atoms within their plane, but are also affected by out-of-plane scattering between atoms in the top two layers. This causes the SFP peak position from the third Se layer to be slightly different than the SFP from the topmost Se layer, which leads to a weakly resolved double SFP in the region from approximately 16° to 30° in Fig. 6.

Despite the complications in understanding the ICISS features, however, the fact that the positions and shapes of all the features of the surface prepared by IBA with a 510°C annealing temperature and the *in situ* cleaved surface match each other indicates that their terminations and surface atomic structures are the same. This further confirms the conclusion reached in Ref. [11] that IBA can prepare surfaces that are as well ordered and have the same structure as those prepared by *in situ* cleaving.

Figure 6 also shows that when the annealing temperature increases from 510°C to 520°C, however, a new feature appears at polar angles in the range of about 5° to 10°. A new feature at a smaller polar angle suggests that annealing above 510°C induces Se vacancies on the surface. The inset to Fig. 6 illustrates how an absent Se atom causes the distance between the two nearest neighbor surface atoms to increase so that the critical angle decreases and the SFP moves to a smaller polar angle [34].

Kalypso simulations are performed to verify that Se vacancies can create such a new feature, with the results shown in Fig. 8. A one-dimensional Se chain with an interatomic distance of 4.14 Å is used to simplify the simulations. Six different target models are employed with different percentages of surface Se vacancies: 0%, 5.6%, 10%, 16.7%, 25% and 50%. The Se vacancies are modeled by removing a fraction of the surface Se atoms periodically. The simulations clearly show a new peak emerging at around 5°-15° that is close to the position of the new feature in Fig. 6 for the 520°C annealed sample, and the intensity of this peak increases with the fraction of Se vacancies. Thus, the simulated polar angle scans support the supposition that the new feature results from Se vacancies created by the high annealing temperature. Note that these vacancies can be located in either or both the first or third Se layers because of the two inequivalent (300) planes that are terminated by Se.

The simulations indicate that the polar angle scans are only moderately sensitive to Se vacancies as a concentration of 5.6% shows just a minimal feature around 10°. Thus, the absence of such a feature in the experimental data of Fig. 6 implies that the concentration of Se vacancy defects is well below 5% on the *in situ* cleaved surfaces and on surfaces prepared by IBA with an annealing temperature of 510°C or less. Furthermore, the experimental observation of the new feature after annealing at 520°C implies that the concentration of Se vacancies formed on the surface after annealing at 520°C is large enough to monitor with ICISS. A rough comparison of the experimental data in Fig. 6 to the simulations in Fig. 8 would suggest a vacancy concentration in the range of 10-15%.

LEIS measurements of the sample holder provide additional evidence for Se removal from the sample by annealing. Figure 9 shows a set of energy spectra collected with the ESA from the Ta sample holder after IBA cycles using different annealing temperatures. The peak at around 0.8 keV is the Se SSP, while the peak at around 1.7 keV is the Ta SSP. Figure 9 shows that Ta sample holder was free of Se after sputtering and when the annealing temperature was less than 510°C (bottom spectrum), but that a Se SSP emerges when the annealing temperature exceeds 540°C. Note that after many annealing cycles, the chamber windows opposite to the sample become covered with a material, presumably Se, suggesting that at least some of the Se desorbs from the $Bi_2Se_3$ surface into vacuum. Although Se could desorb from the sample, scatter from the chamber walls and then land on the holder next to the sample, which would explain the results shown in Fig. 9, this is an unlikely scenario. More feasible is that some of the Se also leaves the $Bi_2Se_3$ surface by lateral diffusion from the sample to the sample holder, but more work is needed to explore the specific mechanisms responsible for the formation of the Se vacancies.

In addition, the IBA-prepared samples are particularly robust to reaction with atmospheric contaminants. Figure 10 shows a TOF-LEIS spectrum collected from an *ex situ* cleaved surface as well as spectra collected from an *in situ* cleaved surface and an IBA-prepared surface that had been exposed to air. The IBA-prepared and *in situ* cleaved samples were removed from vacuum, exposed to air for about two hours and then reinserted into the UHV chamber. The TOF-LEIS spectra collected from these samples show that all of the surfaces remain Se-terminated as the Se SSP is still present and a Bi SSP does not appear. In addition to SSPs, however, some LEIS spectra show features that result from direct recoiling of light adsorbates. Direct recoiling is the removal of fast atoms from a surface by direct collision with the projectile, and it is particularly sensitive to light adsorbates [21]. For $Bi_2Se_3$, direct recoiling has been observed at shorter flight times than the Bi SSP in Ref. [11], and these are indicated in Fig. 10. Direct recoiling of surface contaminants for the air-exposed IBA-prepared surface is barely visible in the data which implies that the concentration of contaminants is much smaller than on an air-exposed cleaved surfaces. This suggests that IBA-prepared surfaces are more resistant to surface chemical reaction than are cleaved surfaces, whether the cleaving is *in situ* or *ex situ*. In Ref. [11], it was proposed that surface adsorption and subsequent chemical reaction was greatly increased in the presence atomic-scale defects, such as those induced by mechanical cleaving. The present results suggest that IBA as a surface preparation method produces materials with fewer of these reactive defects. The fact that AFM images show a higher density of steps on IBA-prepared surfaces than on cleaved surfaces suggests that the steps are not the chemically reactive sites. Instead, atomic scale defects are the likely nucleation sites at which contaminants adsorb and subsequently react.

**IV. Conclusions**

After a single crystal $Bi_2Se_{3.12}$ sample is initially degassed, one 30 min ion bombardment and 30 min annealing cycle is sufficient to form a well-ordered surface that has the same atomic structure as an *in situ* cleaved surface. As it was previously shown that $Bi_2Se_3$ surfaces cleaved *ex situ* have a non-reproducible termination [11], the work presented here demonstrate that IBA can be used to restore such surfaces to a well-ordered Se-termination.

The optimal annealing temperature range for IBA is between 400°C and 510°C. If the temperature is below this range, the surface won't crystallize well enough to produce a sharp LEED pattern. If the annealing temperature is above this range, then Se vacancies will form even though the sample still displays a reasonably sharp LEED pattern. The vacancies are indicated by a new ICISS polar scan feature at ~5°-10° and the emergence of Se on the sample holder. Whether annealing above 510°C causes Se atoms to desorb or diffuse from the sample is still an open question.

The ion scattering methodology used here shows that ICISS can directly measure surface vacancies. This methodology will be improved in subsequent investigations by collecting data using different projectiles and performing simulations with improved statistics. For example, $Li^+$ has a lighter mass and therefore smaller shadow cones than $Na^+$, while $He^+$ neutralizes through an Auger-type mechanism that makes it even more surface sensitive [21]. Thus, a comparison of data collected with these projectiles would make it possible to clearly differentiate between first and third layer Se vacancies. Such information will be of great utility in investigations that correlate the electronic and structural properties of TI materials.

IBA-prepared surfaces are also more resistant to air contamination than cleaved surfaces and are thus likely to be more free of defects. Therefore, IBA preparation of single crystal $Bi_2Se_3$ surfaces is a recommended method for sample preparation for UHV surface and other investigations of this class of materials.

**V. Acknowledgements**

This material is based on work supported by, or in part by, the U.S. Army Research Laboratory and the U.S. Army Research Office under Grant No. 63852-PH-H. CMV was supported by MacREU, a National Science Foundation Research Experience for Undergraduate students site funded under NSF DMR 1359136.

# References

1.  Y. S. Hor, A. Richardella, P. Roushan, Y. Xia, J. G. Checkelsky, A. Yazdani, M. Z. Hasan, N. P. Ong, and R. J. Cava, Phys. Rev. B **79**, 195208 (2009).

2.  S. K. Mishra, S. Satpathy, and O. Jepsen, J. Phys.: Condens. Matter **9**, 461 (1997).

3.  H. Zhang, C.-X. Liu, X.-L. Qi, X. Dai, Z. Fang, and S.-C. Zhang, Nat. Phys. **5**, 438 (2009).

4.  Y. Xia, D. Qian, D. Hsieh, L. Wray, A. Pal, H. Lin, A. Bansil, D. Grauer, Y. S. Hor, R. J. Cava, and M. Z. Hasan, Nat. Phys. **5**, 398 (2009).

5.  X.-L. Qi and S.-C. Zhang, Rev. Mod. Phys. **83**, 1057 (2011).

6.  C. Nayak, S. H. Simon, A. Stern, M. Freedman, and S. Das Sarma, Rev. Mod. Phys. **80**, 1083 (2008).

7.  I. Žutić, J. Fabian, and S. Das Sarma, Rev. Mod. Phys. **76**, 323 (2004).

8.  Z. Jiang, C.-Z. Chang, M. R. Masir, C. Tang, Y. Xu, J. S. Moodera, A. H. MacDonald, and J. Shi, Nat. Commun. **7**, 11458 (2016).

9.  R. W. G. Wyckoff, *Crystal Structures - Volume 2 : Inorganic Compounds $RX_n$, $R_nMX_2$, $R_nMX_3$* (Interscience Publishers, New York, 1964).

10. R. D. Gann, J. Wen, Z. Xu, G. D. Gu, and J. A. Yarmoff, Phys. Rev. B **84**, 165411 (2011).

11. W. Zhou, H. Zhu, and J. A. Yarmoff, Phys. Rev. B **94**, 195408 (2016).

12. A. S. Hewitt, J. Wang, J. Boltersdorf, P. A. Maggard, and D. B. Dougherty, J. Vac. Sci. Technol. B **32**, 04E103 (2014).

13. M. T. Edmonds, J. T. Hellerstedt, A. Tadich, A. Schenk, K. M. O'Donnell, J. Tosado,

N. P. Butch, P. Syers, J. Paglione, and M. S. Fuhrer, J. Phys. Chem. C **118**, 20413 (2014).

14. D. Kong, J. J. Cha, K. Lai, H. Peng, J. G. Analytis, S. Meister, Y. Chen, H.-J. Zhang, I. R. Fisher, Z.-X. Shen, and Y. Cui, ACS Nano **5**, 4698 (2011).

15. V. V. Atuchin, V. A. Golyashov, K. A. Kokh, I. V. Korolkov, A. S. Kozhukhov, V. N. Kruchinin, S. V. Makarenko, L. D. Pokrovsky, I. P. Prosvirin, K. N. Romanyuk, and O. E. Tereshchenko, Cryst. Growth Des. **11**, 5507 (2011).

16. X. He, W. Zhou, Z. Y. Wang, Y. N. Zhang, J. Shi, R. Q. Wu, and J. A. Yarmoff, Phys. Rev. Lett. **110**, 156101 (2013).

17. H. E. Farnsworth, R. E. Schlier, T. H. George, and R. M. Burger, J. Appl. Phys. **26**, 252 (1955).

18. D. Haneman, Phys. Rev. **119**, 563 (1960).

19. P. M. Coelho, G. A. S. Ribeiro, A. Malachias, V. L. Pimentel, W. S. Silva, D. D. Reis, M. S. C. Mazzoni, and R. Magalhães-Paniago, Nano Lett. **13**, 4517 (2013).

20. A. Cavallin, V. Sevriuk, K. N. Fischer, S. Manna, S. Ouazi, M. Ellguth, C. Tusche, H. L. Meyerheim, D. Sander, and J. Kirschner, Surf. Sci. **646**, 72 (2016).

21. W. J. Rabalais, *Principles and applications of ion scattering spectrometry : surface chemical and structural analysis* (Wiley, New York, 2003).

22. J. Dai, D. West, X. Wang, Y. Wang, D. Kwok, S. W. Cheong, S. B. Zhang, and W. Wu, Phys. Rev. Lett. **117**, 106401 (2016).

23. J. G. Analytis, J.-H. Chu, Y. Chen, F. Corredor, R. D. McDonald, Z. X. Shen, and I. R. Fisher, Phys. Rev. B **81**, 205407 (2010).

24. Z. Wang, T. Lin, P. Wei, X. Liu, R. Dumas, K. Liu, and J. Shi, Appl. Phys. Lett. **97**, 042112 (2010).

25. C. B. Weare and J. A. Yarmoff, Surf. Sci. **348**, 359 (1996).

26. M. A. Karolewski, Nucl. Instr. Meth. Phys. Res. B **230**, 402 (2005).

27. W. Zhou, H. Zhu, and J. A. Yarmoff, J. Vac. Sci. Technol. B **34**, 04J108 (2016).

28. D. D. dos Reis, L. Barreto, M. Bianchi, G. A. S. Ribeiro, E. A. Soares, W. S. Silva, V. E. de Carvalho, J. Rawle, M. Hoesch, C. Nicklin, W. P. Fernandes, J. Mi, B. B. Iversen, and P. Hofmann, Phys. Rev. B **88**, 041404 (2013).

29. G. E. Shoemake, J. A. Rayne, and R. W. Ure, Phys. Rev. **185**, 1046 (1969).

30. J. B. Pendry, in Interaction of Atoms and Molecules with Solid Surfaces, edited by V. Bortolani, N. H. March and M. P. Tosi (Springer US, Boston, MA, 1990), p. 201.

31. M. Barat, J. C. Brenot, J. A. Fayeton, and Y. J. Picard, Rev. Sci. Instrum. **71**, 2050 (2000).

32. M. Aono, C. Oshima, S. Zaima, S. Otani, and Y. Ishizawa, Jpn. J. Appl. Phys. **20**, L829 (1981).

33. Th. Fauster, Vacuum **38**, 129 (1988).

34. M. Aono, Y. Hou, R. Souda, C. Oshima, S. Otani, and Y. Ishizawa, Phys. Rev. Lett. **50**, 1293 (1983).

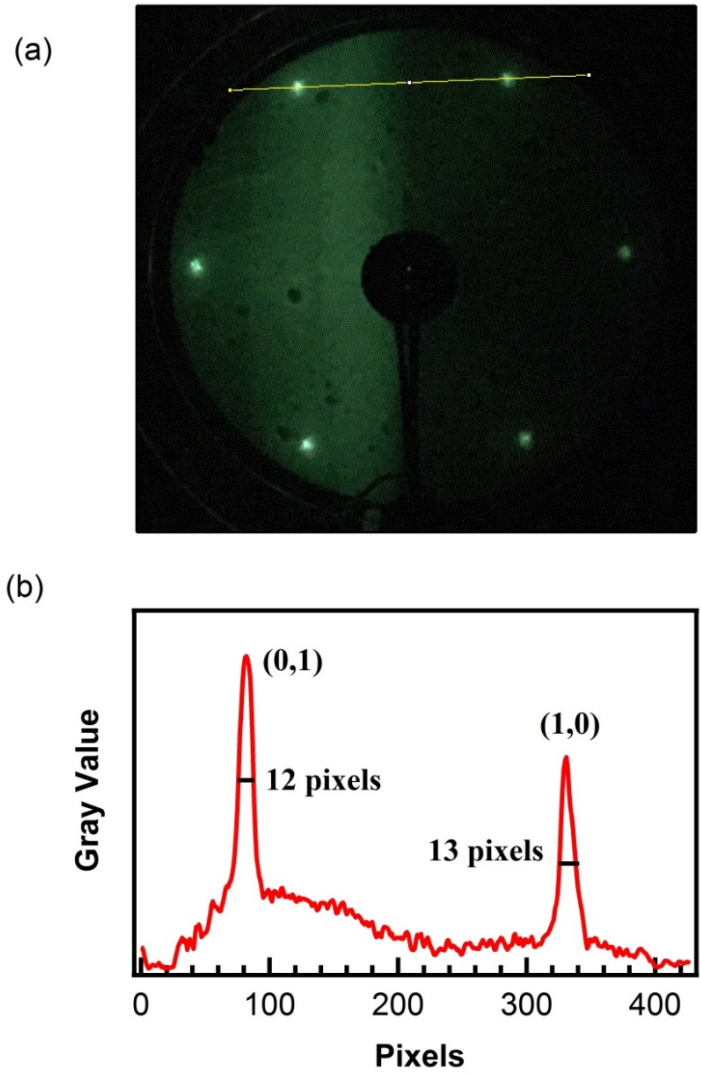

**Figure 1.** (a) A representative LEED pattern collected with an electron beam energy of 25.9 eV from a $Bi_2Se_{3.12}$ surface prepared by IBA using an annealing temperature of 510°C. (b) A gray scale profile along the line indicated in (a), with the FWHM's for LEED spots (0,1) and (1,0) indicated.

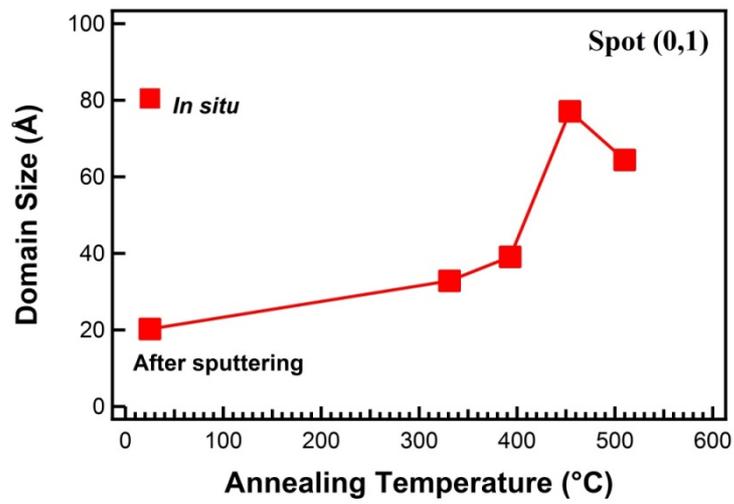

**Figure 2.** Calculated domain sizes of IBA-prepared $Bi_2Se_3$.[12] as a function of annealing temperature. The domain sizes of an *in situ* cleaved surface and a surface after 1 keV $Ar^+$ bombardment (sputtering) for 30 min, but prior to annealing, are included for comparison.

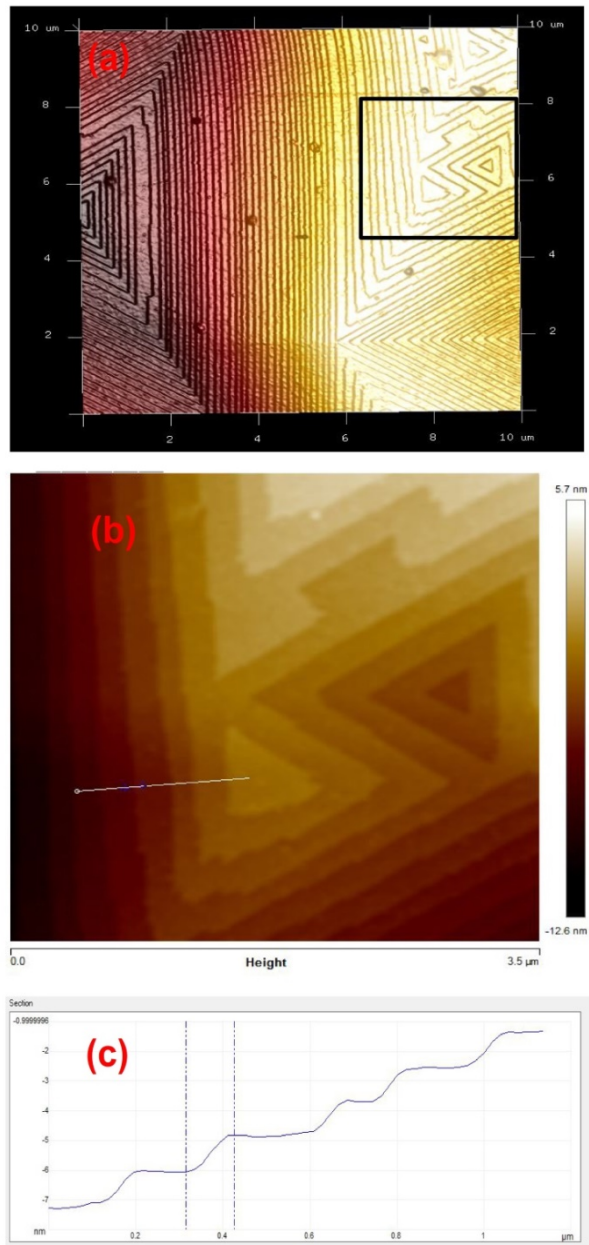

**Figure 3**. (a) An AFM image of a $Bi_2Se_3$ surface prepared by IBA with annealing at 490°C shown in an oblique 3D perspective. (b) A close-up of the boxed area in (a) shown as a 2D image with the indicated color scale representing height. (c) The height profile measured along the line indicated in (b).

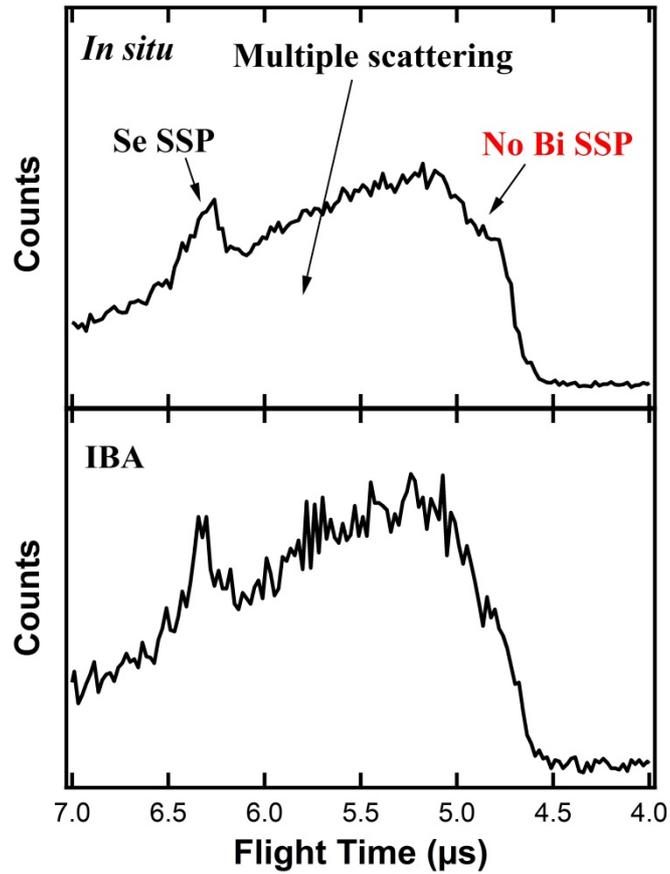

**Figure 4.** TOF spectra collected at a 126° scattering angle along the [210] azimuth employing normally incident 3 keV Na$^+$ ions. The samples are Bi$_2$Se$_{3.12}$ prepared by *in situ* cleaving (upper panel) and IBA with a 510°C annealing temperature (lower panel).

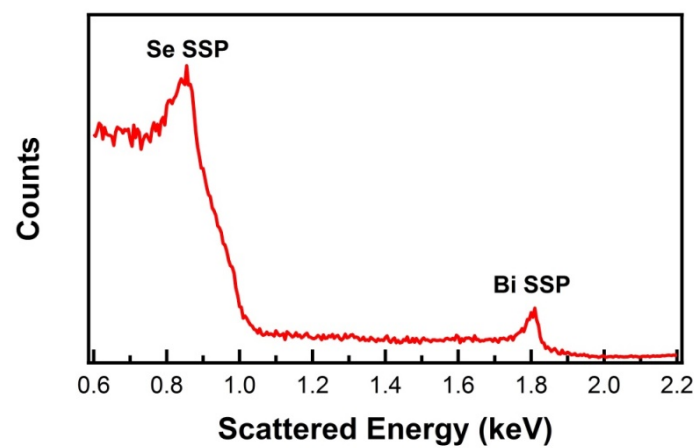

**Figure 5**. Energy spectrum collected with the ESA using 3 keV Na$^+$ scattering from IBA-prepared Bi$_2$Se$_{3.12}$ surface with incident polar angle of 57º along [010] azimuth and a scattering angle of 161º. The sample was prepared using an annealing temperature of 510°C.

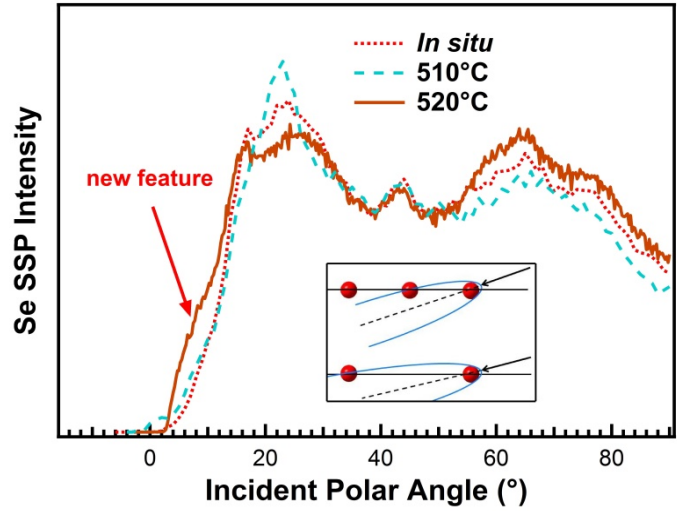

**Figure 6.** ICISS polar scans of the Se SSP using 3 keV $Na^+$ scattering from $Bi_2Se_3$,[12] along the [010] azimuth for *in situ* cleaved surface (short dashed line) and IBA-prepared surfaces with annealing temperatures of 510°C (long dashed line) and 520°C (solid line). Inset: A schematic illustration of the surface Se atoms with the incident ion beam orientated along the critical angle. The upper drawing shows a complete chain of Se atoms, while the lower drawing illustrates a chain with a Se vacancy.

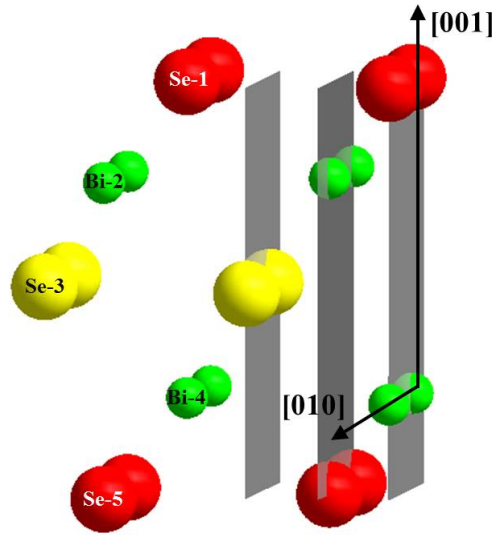

**Figure 7.** A schematic diagram of the outermost QL of $Bi_2Se_3$(001) indicating the three inequivalent (300) planes along the [010] azimuth. The atoms are labeled with their chemical symbol and the atomic layer in which they reside within the QL. For example, "Se-1" indicates the first layer Se in the QL. The sizes of the atoms are shown as their ionic radii. Three of the planes are highlighted by rectangular vertical strips.

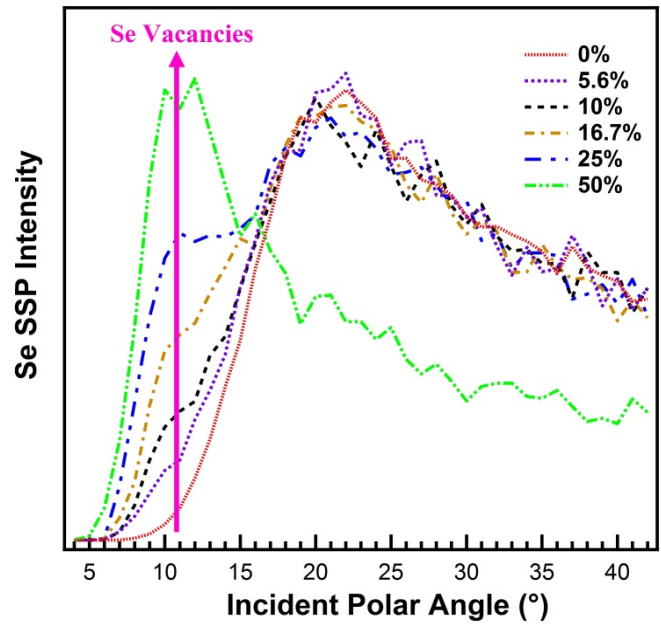

**Figure 8.** Calculated ICISS polar scans of the Se SSP intensity collected along the [010] azimuth using 3 keV $Na^+$ scattering from $Bi_2Se_3$ surfaces with different percentages of Se vacancies.

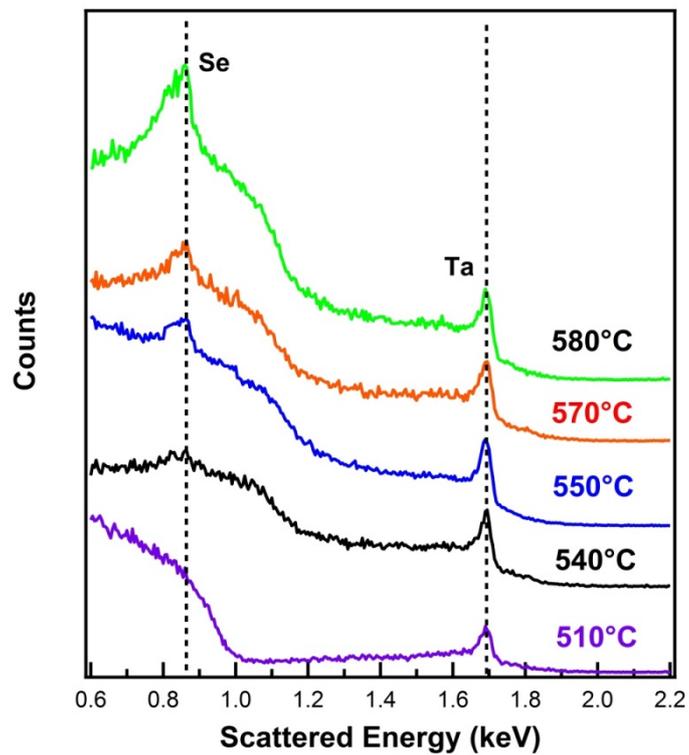

**Figure 9**. Energy spectra collected with the ESA for 3 keV Na$^+$ scattered at a 161° angle from the same spot on the Ta sample holder after the Bi$_2$Se$_3$ samples were annealed at different temperatures.

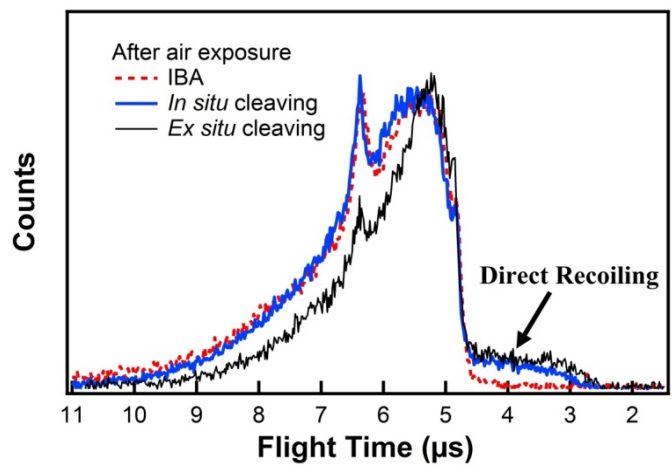

**Figure 10**. TOF spectra employing normally incident 3 keV Na$^+$ ions collected at a 126° scattering angle along the [210] azimuth from an IBA-prepared sample after 140 mins' air exposure (dashed line), an *in situ* cleaved sample after 158 mins' air exposure (thick line), and an *ex situ* cleaved sample with less than 5 mins' air exposure (thin line). The feature due to direct recoiling of light surface contaminants is marked.